\documentstyle[prb,aps,epsf, preprint]{revtex}
\tightenlines

\begin{document}

\title{\bf Nonlinear effects in E$\otimes(b_1+b_2)$ Jahn-Teller model:
Variational approach with excited phonon states and mode correlations.}
\author{ Eva Majern\'{\i}kov\'a$
{}^{\dag \ddag *}$,
S. Shpyrko${}^{\dag **}$ }
\address{${}^{\dag}$Department of Theoretical Physics, Palack\'y University, \\
T\v r. 17. listopadu 50, CZ-77207 Olomouc, Czech Republic \\
${}^{\ddag}$Institute of Physics, Slovak Academy of Sciences, \\
D\'ubravsk\'a cesta, SK-84 228 Bratislava, Slovak Republic
}


\maketitle

\begin{abstract}
Interplay of nonlinear and quantum effects in the ground state of the
 E$\otimes (b_1+b_2)$ Jahn-Teller model was investigated
by the {\it variational approach and exact numerical simulations}.
They result in the recognition of
(i) importance of the admixture of
{\it the first excited state of the displaced harmonic oscillator} of
the symmetric phonon mode in the ground state of the system in the
selftrapping-dominated regime; (ii) existence of {\it the region of
localized $b_1$-undisplaced oscillator states} in the tunneling-dominated
regime. The effect (i) occurs owing to significant decrease
of the ground state energy on account of the overlapping contribution
 of the symmetric phonon mode between the states of
the same parity. This contribution considerably improves variational
results especially in the selftrapping-dominated regime.
Close to the E$\otimes$e limit,
the nonlinear effects of {\it two-mode correlations}
turn to be effective due to the rotational symmetry of this case.
In the tunneling-dominated regime the
phonon wave functions behave like the strongly localized harmonic
oscillator ground state and the effect (i) looses its significance.
\end{abstract}

pacs[63.20.Kr,31.30.Gs,71.70.Ej ]

\section{Introduction}

Recently, revival of interest in two-level electron-phonon systems
occurs owing to the experimental evidence that Jahn-Teller
structural phase transition occurs in spatially anisotropic complex
structures (fullerides, manganite perovskites, etc)
\cite{Kaplan:1995}$^{,}$\cite{Gunnarson:1995}$^{,}$\cite{Muller:1999}$
^{,}$\cite{Obrien:1993}$^{,}$\cite{Dunn:2002}.
 
 JT model is a protopype model for phonon removing the degeneracy of electron
 levels \cite{Obrien:1993}$^{,}$\cite{Kaplan:1995}.
Current investigations were focused mainly on E$\otimes$e JT model
with electron coupling to two degenerate
intramolecular phonon modes, an antisymmetric
and symmetric  one with respect to the reflection.

The reflection symmetry of two-level electron-phonon models like
the exciton and the dimer model with
onsite electron coupling to one phonon mode implies
nonlinear peculiarities of quantum nature \cite{Wagner:1989}.
For these models, Shore et al. \cite{Sander:1973} introduced
variational wave function in a form of linear combination of the
harmonic oscillator wave functions related to two levels of different
parity with respect to the reflection. This picture can be understood in terms
of two or more asymmetric local minima of the effective polaron potential
(i.e. the potential energy expression for the trial wavefunction in the 
space of variational parameters). 
Here,
respective ground state wave function can be approximated by a linear
combination of two oscillators with parameters corresponding to these local 
minima and
coupled by means of further variational parameters.
This approach was shown to yield the
lowest ground state energy for the two-level models
\cite{Sander:1973}$^{,}$\cite{Wagner:1994}.
The peculiarities due to reflection phenomena occurred in the rotation-symmetric E$\otimes$e JT model
\cite{Wagner:1992} as well. Strongly localized
 non-displaced phonon  (exotic) states appeared in the numerical spectra.
 They were considered useful for interpretation of the "fast" component of
 luminiscence spectra.
However, in order to respect the rotational symmetry of the $E\otimes e$
model the proper variational approach should be formulated in radial coordinates
\cite{Barentzen:2001}.

 In crystals exhibiting high spatial anisotropy with tensor properties
 of bulk characteristics (e.g. perovskites, fullerids, etc.)
 the rotation symmetry of Jahn-Teller molecules is generally broken.
Therefore, it is reasonable to investigate JT model assuming
different coupling strengths $\alpha$ and $\beta$ for the onsite intralevel
and interlevel electron-phonon couplings, respectively
(E$\otimes(b_1+b_2)$ model \cite{Kaplan:1995}).
Such a model can be also considered as a generalization of the
exciton-phonon or the dimer-phonon model assuming the electron tunneling
 to be phonon-assisted.

In order to understand the physical nature of the
nonlinear effects we propose variational ansatz inspired by the
shape of the numerical ground state wave function (Fig.1):
the ``principal'' part of
Gaussian character for both oscillators (in the absolute minimum of the
nonlinear effective potential (Fig.1 of our previous paper
\cite{Majernikova:2002}
 and Fig.2 of this paper) and a minor ``reflective'' part which corresponds to
the another (local) minimum of the potential.
For this minor part we consider 
the admixture of the first excited harmonic
oscillator of the symmetric mode (rather than the only ground oscillator state,
as it was commonly considered elsewhere). This admixture in the variation 
trial function leads to essential improvement of the
results, as it will be shown in Sect. III.

 Formulation of the variational ansatz 
and calculation of the ground state energy is presented in the Section II.
In the Section III, analysis of the interplay of quantum effects and nonlinearity
and related discussions as well as the reliability of different
variational alternatives was investigated by
comparison with results of exact numerical simulations.

\section{Variational wave function of the generalized
 JT model }

We investigate local spinless double degenerate electron
states linearly coupled to two intramolecular phonon modes
described by Hamiltonian

\begin{equation}
H= \Omega  (b_{1}^{\dag}b_{1} +b_{2}^{\dag}b_{2}+1 )I +
\alpha  (b_{1}^{\dag}+b_{1})\sigma_{z}
 -\beta (b_{2}^{\dag}+b_{2})\sigma_{x},
 \label{1}
\end{equation}

where $\sigma_x=\left (\matrix {0,\ 1\cr 1, \ 0}\right )$,
$\sigma_z=\left (\matrix {1,\ 0\cr 0,  -1}\right )$ are Pauli matrices,
$I$ is the unit matrix.
This pseudospin notation refers to two-level electronic system, the Hamiltonian  
is thus $2\times 2$ matrix.

 The antisymmetric phonon mode $b_{1}$ splits the degenerate unperturbed electron level
 ($j=1,2$) while the symmetric mode $b_{2}$ mediates the electron
transitions between the levels. This latter term represents phonon-assisted
tunneling, a mechanism of the nonclassical (nonadiabatic) nature as well as
 is the pure tunneling in related exciton and dimer models. Evidently,
for $\beta=0$, the one-level Holstein model (\ref{1}) is harmonic, while
the coupling to the higher level ($\beta \neq 0$) is the origin of the
strong nonlinearity in the phonon space as will be seen below.

For  $\beta=\alpha $, the  interaction part of (\ref{1})
yields the rotationally symmetric $E\otimes e$ form  \cite{Obrien:1993}
with a pair (an antisymmetric and a symmetric under reflection) of
double degenerated vibrations.

The general case $\alpha\neq\beta$ exploited here breaks the 
common invariance of $E\otimes e$ Jahn-Teller model under the exchange of
$1$ and $2$ phonons; in other words, the rotational symmetry of the 
Hamiltonian (\ref{1}) is broken while the
reflection symmetry is kept.
This symmetry is inherent property of the Jahn-Teller model and it is
crucial for the diagonalization of (\ref{1}) and construction of the
variational ansatz as it is evident from what follows.

Hamiltonian (\ref{1}) can be diagonalized in the electronic subspace
 using the Fulton-Gouterman  unitary operator
\cite{Fulton:1961}
\begin{equation}
 U= \frac{1}{\sqrt 2} \left ( \matrix{1\ , \ G_{1}\cr  1\ ,
\ -G_{1}}\right ), \quad G_{1} =\exp (i\pi
b_{1}^{\dag}b_{1}),
\label{2}
\end{equation}
 as  follows
 \begin{eqnarray}
 H_{FG} =  U H U^{-1}=
\Omega
\left (b_{1}^{\dag}b_{1} +b_{2}^{\dag}b_{2}+1\right ) 
+\alpha (b_{1}^{\dag}+b_{1}) I
-\beta (b_{2}^{\dag}+b_{2})\sigma_z G_{1} \equiv
H_{ph}+H_{\alpha}+H_{\beta}.
\label{3}
\end{eqnarray}

The operator $G_1$ in (\ref{2}) is the phonon reflection operator:
$G_1 (b_1^{\dag}+b_1)=-(b_1^{\dag}+b_1), \ G_1^2=1$.
We see that Fulton-Gouterman transformation reveals high
nonlinearity in the system (term with $G_1$ in (\ref{3})),
otherwise this nonlinearity was
hidden in the initial Hamiltonian (\ref{1}).
Hamiltonian 
distincts from the exciton (dimer) by the phonon-2
assistance of the tunneling amplitude $\beta G_1$.  The factor $G_1$
(\ref{2}) represents continuous virtual emission and absorption of the
phonons $1$ and  mediates Rabi oscillations of the electron between the levels.
 These quantum oscillations are essentially the origin of the
 nonlinearity of the reflection symmetric model as will be seen below.

The full reflection operator is $G=G^{el}G_{1}$, where
the electron reflection operator is defined by $G^{el}|1\rangle=|2\rangle$.  
Equivalently to the FG transformation
one can exploit commutation of $G$ with Hamiltonian (\ref{1}),
 $ [H, G]= 0$, so that the wave function of Hamiltonian (\ref{1})
 related to the representation of the
 inversion group $p=\pm 1$ is a linear combination of the base functions
 \begin{equation}
|\Psi^{(p)}\rangle = \frac{1}{\sqrt 2}(1+p G) |1\rangle |
\phi^{(p)}\rangle,
\label{4}
\end{equation}
where $G_{1} |\phi^{(p)}\rangle =|\phi^{(-p)}\rangle $.
Hamiltonian (\ref{3}) though diagonalized is no more reflection symmetric,
but the interaction part is antisymmetric against reflection. As a
consequence, in the limit $\alpha=\beta$  the rotation symmetry is
broken. Therefore, use of the FG transformation in the case of
E$\otimes$e JT is inappropriate: it breaks the symmetry which is
necessary for the proper choice of the ground state in this case
(see the discussion in the Conclusion).

Inserting the representation of the wave functions (\ref{5})
into the Schr\"odinger equation  related to (\ref{1}) we are left with
the Fulton-Gouterman equation
\begin{equation}
H_{FG}^{(p)} \phi ^{(p)}  =
[\Omega\left (b_{1}^{\dag}b_{1} +b_{2}^{\dag}b_{2}
+1\right ) +\alpha (b_{1}^{\dag}+b_{1})  - p
\beta (b_{2}^{\dag}+b_{2}) G_{1}) ]\phi^{(p)} =E^{(p)}\phi^{(p)}, \ p= \pm 1.
\label{5}
\end{equation}
It is evident that procedure yielding the set (\ref{5}) is equivalent to the FG
transformation yielding  Hamiltonian (\ref{3}) in the Pauli
 $2\times 2$ matrix representation.
From the diagonalized form (\ref{5}) we can see that $\beta$-coupling
breaks the degeneration of two electron levels which do not have the same
energy.

As it was pointed out to us,  Hamiltonian
(\ref{1}) and, equivalently also (\ref{5}) (and all the related
quantities), are symmetric
against simultaneous exchange of $p\leftrightarrow -p$,
$\alpha\leftrightarrow -\beta $ and,
ensuingly, in terms of parameters introduced later on,
$\chi=\beta/\alpha\leftrightarrow
1/\chi $ and $\mu\equiv \alpha^2/\Omega^2\leftrightarrow
\beta^2/ \Omega^2$.

In what follows we shall investigate variationally
{\it the ground state ($p=1$)} with lower energy only in
the representation of Eq. (\ref{5}).

 In two-level electron-phonon systems with linear coupling the
coherent phonon subsystem does not conserve the number of phonons. Therefore,
 the upper level will share partly the distribution of
phonons even in the ground state.
From this reason, the two-center wave function in the form of
asymmetric nonunitary ansatz with a variational parameter $\eta$
proposed by Shore et al \cite{Sander:1973} and Sonnek et al
\cite{Wagner:1994}  for exciton or dimer models coupled to one
phonon mode
\begin{equation}
|\Psi ^{(1)}\rangle=  \frac{1}{\sqrt  C_0}(1+\eta  G)|1\rangle
|\phi^{(1)}\rangle ,
\label{6}
\end{equation}
($C_0$ is a normalization constant)
was proved to yield better (lower) estimation of energy of the ground state
\cite{Sander:1973}$^{,}$\cite{Wagner:1994} when compared with the
eigenfunction $|\phi^{(1)}\rangle $ of $H_{FG}$ (\ref{5}).

Our present suggestion of the variational ansatz for the ground state is motivated by
the numerical solution for the wave function to the diagonalized Eq.
(\ref{5}).
In the following numerical and analytical calculations it is convenient
to use two basic parameters: the asymmetry parameter $\chi=\beta/\alpha$ and
the effective coupling strength $\mu=\alpha^2/2\Omega^2$.
The wavefunctions for the strong coupling $\mu=2$ and $\chi=0.9$, $\chi=1$, 
and $\chi=1.5$
are given in Figs. 1a,b,c
(the standard numerical simulation procedure is outlined
in the next Chapter) where the wavefunctions are depicted in the coordinate 
representation in the space of two phonon oscillators $Q_1\otimes Q_2$.
Three distinct forms of the phonon solutions correspond to small, intermediate and 
large values of the parameter $\chi$.

Fig. 1a represents the ``selftrapping'' region in which contribution of
the electron transitions between the levels (assisted by the oscillator
2) is small. Here the main Gaussian of the wave function $\phi^{+}(Q_1)$
at negative values of $Q_1$
($p=+1$)  is accompanied in the $Q_1$-subspace by the
reflective part $\eta\phi^{-}(-Q_1)$ corresponding to $p=-1$ parity which have already inspired the
introduction of the nonunitary ansatz  (\ref{6}) representing
the minor reflection respective to the
axis $Q_1=0$ \cite{Sander:1973}. However, 
in Fig. 1a, the admixture of the first excited state of the
oscillator 2 (coordinate $Q_2\approx 0$)
of the parity $p=-1$ (displaced to the right to $Q_1>0$) can be recognized,
while for $p=+1$
both oscillators (region $Q_1<0 $) remain in the Gaussian ground state.
The  variational treatment which is the topic of present paper aims mostly 
at capturing the situation of Fig. 1a as adequate as possible.

In Fig. 1b ($\chi=1$) which represents the case of E$\otimes$e Jahn-Teller
molecule, the rotational symmetric nature of the ground state at
$\chi=1$ is easily recognizable. We mention by passing that an ideal rotation 
symmetry of the phonon wavefunction (``mexican hat''\cite{Obrien:1993})
which might be expected in the
adiabatic case (that is that of big $\mu$) of a rotationally symmetric problem can 
not be reached even for very large 
$\mu$, as it was explained in detail by Eiermann et al\cite{Wagner:1992}.
For the symmetric $E\otimes e$ Jahn-Teller case we have merely an approximation to the
``mexican hat'' wavefunction profile spoiled by the non-zero angular momentum 
part which prevents the profile to show the complete rotational symmetry. This 
picture combines in it the features of self-trapping (Fig. 1a) and
tunnelling (Fig. 1c) cases.
The region close to E$\otimes$e Jahn-Teller case appears also to show
the most serious discrepancies of the variational treatment.

Fig. 1c represents another limit case - that of large values of $\chi$
corresponding
to the (quantum) region of dominated phonon-assisted tunnelling. 
There the Gaussian form
of the wave function is retained, although the second oscillator
is displaced towards $Q_2>0$, while {\it phonons 1 remain undisplaced}
(the Gaussian is centered at $Q_1\approx 0$). 

In Figs. 2a,b,c, we sketched three  shapes of the effective potential
(energy expression from the trial functions)
controlled by the displacement parameters $\gamma_1$ and $\gamma_2$ only which 
refer to the variants of the wave
functions in Figs. 1a,b,c, respectively.

Thus, guided by Fig. 1a,
the variational wave function can be proposed in the following nonunitary form
\begin{equation}
\Psi= \frac{1}{\sqrt C}\left[\phi_0^{(+)}(\gamma_1,\gamma_2,r_1,r_2, \lambda)
+\eta_1 \phi_0^{(-)}(-\gamma_1,\gamma_2,r_1,r_2, -\lambda)+\eta_2 \phi_1^{(-)}
(-\gamma_1,\gamma_2,r_1,r_2, -\lambda)\right ]
\label{7}
\end{equation}
where $7$ variational parameters $\gamma_i$, $r_i$, $\eta_i$, and $\lambda$
 are introduced and defined below.

Phonon wave functions $\phi_i^{(\pm)}, \ i=0,1$, are supposed to be squeezed
coherent and correlated oscillators produced applying the set of generators 
on the phonon vacuum state
as follows:
\begin{eqnarray}
\phi_0^{(\pm)}(\gamma_1,\gamma_2, r_1,r_2,\lambda)=
D_1(\pm \gamma _1) S_1(r_1) D_2(\gamma _2)S_2(r_2)S_{12}
(\pm\lambda)|0\rangle,\nonumber\\
\phi_1^{(-)} (\gamma_1,\gamma_2, r_1,r_2,\lambda)=  D_1(- \gamma _1) S_1(r_1) D_2(\gamma _2)S_2(r_2)S_{12}
(-\lambda)b_2^{\dag}|0\rangle,
\label{8}
\end{eqnarray}
($|0\rangle$ is the phonon-$1,2$ vacuum state; indices $0,1$ at $\phi_0, \phi_1$
denote the ground and the first excited state of displaced phonons, 
respectively).

Here we defined the generators of variational displacements $\gamma_i$
\begin{equation}
D_i (\gamma_i)= \exp [\gamma _i(b_i^{\dag}-b_i)],
\label{9}
\end{equation}
and those of squeezings parametrized by $r_i$
\begin{equation}
S_i(r_i)= \exp[r_i(b_i^{\dag 2}-b_i^{2})]
\label{10}
\end{equation}
which are functions of variational parameters of displacement
$\gamma_i$ and squeezing $r_i$ for $i=1,2$.
In Eq. (\ref{5}) the phonon modes $1$ and
$2$ appear coupled in a highly non-linear way (through the term with $\beta$),
therefore one also includes into the ansatz (\ref{7}) the mode correlation
generator
\begin{equation}
 S_{12}(\lambda)= \exp[\lambda(b_1^{\dag}b_2^{\dag}-b_1b_2)]
 \label{11}
 \end{equation}
 with the correlation variational parameter $\lambda$.

The functions $\phi_0^{(+)}$ and $\eta_1\phi_0^{(-)}$ in (\ref{7}-\ref{8})
represent displaced and squeezed  oscillators in $Q_1\times Q_2$ space
whose weight is shifted to the points $(+\gamma_1,+\gamma_2)$ and 
$(-\gamma_1,+\gamma_2)$ respectively (see Fig.1a); The function $\phi_0^{-}$
is merely the reflection of $\phi_0^{(+)}$ weighted by $\eta_1$:
$\phi_0^{(-)}=G_{1}\phi_0^{(+)}$ (see (\ref{4}-\ref{6})).

The function $\phi_1^{(-)}$ is the excited oscillator
$b_2^+|0>_1|0>_2$ displaced likewise $\phi_0^{(-)}$  into the point
$(-\gamma_1, \gamma_2)$ of $Q_1\otimes Q_2$ space and squeezed as well 
(by parameters
$r_1, r_2$) with weighting parameter $\eta_2$. In what follows we show
that introducing this {\it admixture of the excited state of the oscillator-2
in (\ref{7}) essentially improves variational results}.

The last variational parameter $\lambda$ enters in generators
$S_{12}(\pm\lambda)$ which mix phonon modes together; this can be visualized 
as effective rotation of the two-dimensional Gaussian in the plane $(Q_1,Q_2)$;
different signs $\pm\lambda$ keep trace of the reflection symmetry against 
the line $Q_1=0$ (it is best seen from Fig.1b that left and right ``hills'' 
should rotate in the opposite directions).

The complete expression for the mean value of the Hamiltonian
$H=H_{ph}+H_{\alpha}+H_{\beta}$ (\ref{3}) in
the state (\ref{7})-(\ref{11})  is given in Appendix B. A useful 
representation of that expression
can be the following decomposition which separates the contributions of
``ground'' and ``excited'' parts of  trial functions:
\begin{eqnarray}
 \langle H\rangle = \frac{1}{C}\left
 [\langle(\phi_0^{\dag}+\eta_2 \phi_1^{(-)\dag})H
 (\phi_0+ \eta_2\phi_1^{(-)})\rangle\right ] \nonumber\\
= \frac{1}{C}\left [\langle \phi_0 ^{\dag}H \phi_0 \rangle +
2\eta_2 \langle \phi_0^{\dag} H  \phi_1^{(-)\dag}\rangle + \eta_2^2 \langle
 \phi_1^{(-)\dag} H \phi_1^{(-)}\rangle\right ],
 \label{12}
 \end{eqnarray}
 where $\phi_0 =\phi_0^{(+)}+\eta_1 \phi_0^{(-)}$,
 and
 \begin{eqnarray}
C=  \left (1+\eta_1^2+2\eta_1\varepsilon + \eta_2^2\right ), \quad
\varepsilon = \frac{\exp \left (-\frac{2\tilde \gamma _1^2}{\cosh 2\lambda
}\right )}{\cosh 2\lambda}.
\label{B6}
\end{eqnarray}

 The effective Hamiltonian  (\ref{12}) involves a highly nonlinear interplay
of the variational parameters: the admixture of the state $\phi_1^{(-)}$
contributes by the terms due to overlapping of the ground and first excited
state of the oscillator $\propto\eta_2$ and by its own excitation energy
$\propto\eta_2^2$.

In what follows, we investigate joint effects of quantum fluctuations and
nonlinearity in the ground state of (\ref{12})
by minimalization of the energy expression with respect to involved VPs
$\gamma_1, \gamma_2, r_1, r_2, \lambda, \eta_1, \eta_2$.
Parameters of the displacement $\gamma_1, \gamma_2$ are defined by the
displacement generators $D_i (\gamma_i)$ (\ref{9}), parameters of
squeezing $r_i$ by the generators of squeezing $S_i(r_i)$ (\ref{10}),
parameter of the mode correlation $\lambda $ by the generator of the
correlation $S_{12}(\lambda)$ (\ref{11}) and the parameters of
asymmetry $\eta_1, \eta_2$ by the linear combination (\ref{7}).

\section{Interplay of quantum and nonlinear effects in selftrapping and
tunneling}

As it was shown in the last Section, the reflection symmetry
hidden in the original Hamiltonian reveals in Eq. (\ref{5}) due to the
diagonalization
by FG transformation in a highly nonlinear way. This nonlinearity implies
new purely quantum region of the ground state with strong mixing of
the nonlinearity and quantum fluctuations. There are several mechanisms
supporting the quantum (nonadiabatic) fluctuations:

 In the present model the relevant nonadiabaticity parameter
is the ratio of the frequency and the coupling parameter $\Omega/ \alpha$.
The ratio of the polaron energy  $\frac{\alpha^2}{\Omega}$ and of the
frequency $\Omega$,  $\alpha^2/\Omega^2=2\mu $ is a measure of the competition
between the classical (polaron selftrapping) and quantum effects due to
zero energy fluctuations. The quantum effects related to $\Omega$ are
thus relevant at weak couplings $\mu$.

The competition between the selftrapping ($\alpha$) and tunneling
($\beta$) terms results in occurrence of two regions of the ground state:
In the phase plane $\chi $, $\mu$, the ground state
exhibits two phases separated by the crossover line close to $\chi=1$
(Figs. 3, 4, 5,  see also pertaining discussion in our earlier paper
\cite{Majernikova:2002}).
It means that the effective polaron potential exhibits two competing minima
(Fig. 2a) governed by the model parameters $\mu $
and $\chi$. The minima coincide within the border of the regions lying close to
the line $\chi=1$ (Fig. 2b).
The phase $\chi<1$ is selftrapping dominated,
with quantum fluctuations reflected in parameters
$r_1, r_2, \lambda, \eta_i$. The phase $\chi>1$ is the phonon 2-assisted
tunneling dominated region with continuous
virtual emission and absorption of phonons $1$. This {\it phonon-1 exchange
couples the levels within one minimum displaced merely by $\gamma_2$ due to
the phonons 2.}
This minimum is much more sensitive to the
change of model parameters $\mu, \chi $ as well as to quantum fluctuations
reflected in $r_1, r_2 $, $\lambda$ and insensitive with respect to $\eta_i,$ while
$\eta_i\approx 0$.

From the electronic point of view, 
the electron in the selftrapping dominated region is trapped by the phonons
$1$, but due to the interactions mediated by the phonons 2 it can
tunnel to the higher level. Then, owing to the reflection symmetry of the
phonons 2, continuous oscillations of the electron at simultaneous
virtual emission and absorption of phonons 1 occur. These oscillations couple
the levels and thus the electrons into pairs localizing them in one
minimum (Fig. 2c). This mechanism was described
in a recent paper \cite{Majernikova:2002} for a one-dimensional lattice model.

An insight to the importance of different variational parameters can be
gained by analytical minimalization of
Hamiltonian (\ref{12}) in various approximations.
 Numerical considerations show that contributions from the quantum
 parameters $r_i, \eta_i, \lambda$ are at least by an order
 smaller than those from the classical parameters.
 Including also $\eta_2$, we get approximately ($\Omega=1$):
\begin{equation}
\langle H\rangle = \gamma_1^2+\gamma_2^2 +1 +
2\alpha\gamma_1\frac{(1-\eta_2^2)}{(1+\eta_2^2)}
-2\beta\gamma_2 \varepsilon +2\eta_2(\gamma_2\varepsilon-\beta).
\label{12a}
\end{equation}

 Assuming all nonadiabatic parameters
small and minimalizing (\ref{12a}) we get approximately
\begin{equation}
 \gamma _1 \left( 1+ 4\beta^2 \varepsilon^2\right )=
-\alpha \, , \quad
\gamma _2= \beta\varepsilon
 \, ,
\label{13}
\end{equation}
where $\varepsilon\propto \exp (-2\gamma_1^2)$  (\ref{B6}).
From these equations, approximate expressions for both regions (small
and large $\chi$) are summarized below:

In the ``selftrapping'' region $\alpha > \beta$ ($\chi<1$) there is
$\varepsilon \ll 1$ and we get:
\begin{equation}
\gamma_1\simeq -\alpha
 \, ,\quad
\gamma_2= \beta \varepsilon \ll 1,
\label{14}
\end{equation}
(that is small $\gamma_2$ and large negative $\gamma_1$);
In the framework of this approximation (taking also $\eta_2 $ small)
respective ground state energy in the selftrapping region results in
\begin{equation}
E_G^{\alpha}\approx 1-\frac{\alpha^2}{1+8\beta^2\exp (-4\alpha^2)}-
\beta^2 \exp (-4\alpha^2).
\label{14a}
\end{equation}

For the tunnelling dominated region, vice versa:
\begin{equation}
\gamma_1\simeq-\frac{\alpha}{1+4\beta^2}\ll 1 \, , \quad
\varepsilon\simeq 1-2\gamma_1^2
\, , \quad \gamma_2=\beta\varepsilon\simeq \beta.
\label{15}
\end{equation}
In this case the ground state energy is approximated by
\begin{equation}
E_G^{\beta}\approx 1-\frac{\alpha^2}{1+8\beta^2}- \beta^2.
\label{14b}
\end{equation}
Comparing both ground state energies (\ref{14a}) and (\ref{14b})
we observe the asymmetry
against the exchange of the $\alpha$ and $\beta $
of both results due to
 the screening of the tunneling term $\propto \beta^2 $ by
 $\varepsilon=\exp(-2\gamma_1^2)$
 which is either vanishingly small (\ref{14}) or $\propto 1$ (\ref{15}).
Evidently, it is caused by the presence of the nonlinear $G_1$ factor in
the Fulton-Gouterman Hamiltonian (\ref{3}) or (\ref{5}).

Let us examine more thoroughly the parameter $\eta_2$
representing the relative
weight of the ``excited'' admixture. Qualitatively important features brought 
by the Ansatz  (\ref{7}) can be found analytically merely from the
simplified version of the Hamiltonian (\ref{12}) and
(\ref{B2})-(\ref{B5}) with only displacements $\gamma_i$ and $\eta_2$.

The equation for optimized value $\eta_2$ for both regions reads exactly as

\begin{equation}
\eta_2 (1-4\alpha\gamma_1)-(1-\eta_2^2)\beta(1-\varepsilon^2)=0\,;
\label{16}
\end{equation}
Inserting there the above expressions for $\gamma_i$ (\ref{13}-\ref{15}) we get:

\begin{itemize}
\item 
for the selftrapping region 
\begin{equation}
\eta_2^{(I)}\simeq\frac{\beta(1-\varepsilon^2)}{1+4\alpha^2},
\label{17}
\end{equation}
\item
for the tunnelling region
\begin{equation}
\eta_2^{(II)}\simeq \frac{4\beta \gamma_1^2}{1-4\alpha\gamma_1}
\simeq 0\, ,
\label{18}
\end{equation}
\end{itemize}
$\gamma_1 $ given by (\ref{15}).

This analytical estimation shows that the admixture of the excited 
$2$-phonons should play the most important part in the selftrapping region only; 
This conclusion is in the complete accordance with the shapes of the wave functions
for both regions (Figs 1a, 1c),  as well as with further exposed results of
minimalization of variational energies (Figs. 3-5).

Further, exploiting the influence of $\lambda$ and $\eta_1$ separately
(using (\ref{12}), Appendix B and (\ref{14}) for $\chi<1$ ), we get estimations:

\begin{eqnarray}
\sinh(2\lambda)\simeq -2\beta\varepsilon\gamma _1\simeq
4\chi\mu\exp(-4\mu), \quad  \eta_1\simeq-\frac{\beta \gamma _2}{\alpha
\gamma _1}\simeq \chi^2\exp(-4\mu),
\label{19}
\end{eqnarray}
where $ \mu= \alpha^2/2\Omega^2$ was defined at the beginning of the
Section III. as the parameter of the effective interaction.
(The first dependence of (\ref{19})  is recognizable on Fig. 6 and will
be discussed later
in relation to the necessity of accounting for mode correlation).


We calculated then the optimized values of the variational parameters finding 
numerically the minimum of the energy functional in different approximations
 - starting from the most complicated case of complete expression (Appendix B) 
with 7
 parameters included and ending with the adiabatic ansatz with the 
displacements $\gamma_1, \gamma_2$ only.

In order to check the validity of the variational calculations we performed
also the numerical diagonalization of the Hamiltonian in the phonon-$1,2$ space.
We truncated the (infinite) phonon space by $N_1$ 1-phonon states
and $N_2$ 2-phonon states, thus the state vector is $N_1\times N_2$
dimensional. As numerical diagonalization results show,
about 20-50 phonon states are sufficient for convergence.
In Figs. 3 to 5 we also show the results of numerical diagonalization of
Hamiltonian matrix as function of $\chi$ for $\mu=0.5$, $\mu=1$ and $\mu=4$. In the first two
cases we took 20$\times$20 state vector, while in the latter case to
achieve satisfactory convergence
(especially for the tunneling-dominated region when $\chi>1$)
we had to increase the number of phonons-2 up to 50.

The two energy curves from exact numerical solution and ``adiabatic'' 
variational treatment present correspondingly the lower and upper bound for
variationally calculated energy, thus any reasonable variational results
should lie between these bounds, and reliability of a variation ansatz for 
a given parameter region can be judged according to how close the 
corresponding ground 
state energy is to these limits (Figs. 3-5).

We limited ourselves by showing merely some crossections of the plane
$\chi, \mu$ which seem to be typical in discussing the validity
of the variational approach vs ``true'' solution via numerical
diagonalization in the phonon subspace. 
From Figs. 3b to 5b 
showing the differences between ``variational'' and ``exact'' energy for various 
variational approximations
it is seen that while the curves with all
parameters $\eta_i, \lambda, r_i, \gamma _i$ included give minimal discrepancy
(which is evident, since increasing the number of variational
parameters within the same trial function class leads to improving the results), the maximal
discrepancy, and hence the maximal effect of additional
parameters $\eta_i$, $\lambda$ is observed near the line $\chi\simeq 1$,
this region of observed maximal discrepancy being shifted
to the point $\chi = 1$ with growing $\mu$.


From the results of numerical minimalization of the variational expressions for
energy (see Fig. 3-5)
one can see that the ``excitation-reflection'' ansatz (ERA) (\ref{7})
($\eta_2$ included)  results in {\it fascinating improvement of the
variational simulation} of the problem in comparison with the
``simple reflection''  ansatz (SRA) ($\eta_1$ only); especially
it is evident from Figs. 3b, 4b, 5b, where the differences
between variational and ``exact'' energies are plotted.
Although this ansatz can be to a larger extent
inspired by merely the shape of the wave function (Fig. 1a) one might wonder that
the ansatz containing
an excited oscillator as the ``reflective'' part of the trial function lowers
the total energy of the system
in comparison with the ansatz containing the ``zero'' state of the
oscillator ($\eta_2=0$).
An insight to better
understanding this phenomenon can be gained examining the variational energy
expression (\ref{12}), (\ref{B2})-(\ref{B5}).
The energy can be split into three parts 
representing respectively the energy of the ``main'' Gaussian, that of the reflection
part and ``overlapping'' exchange terms
(those containing $\eta_i^0$, $\eta_i^2$, $\eta_i^1$ respectively). Indeed when we
switch from SRA towards ERA the energy
of the reflection part is increased by $\sim \eta_2^2$ by virtue of one
extra displaced ``phonon'' (\ref{B5});
But, if one compares the overlapping
terms for both expressions (below) one can see that it is
{\it the overlapping term of ERA (\ref{B3}), (\ref{20}) which significantly decreases the
overall energy, while the respective overlapping term of SRA (\ref{B2}),
(\ref{21}) contributes only slightly}.

The main contribution to the overlapping integral is contained in the term
  $\ \langle\Phi_0(\gamma_1)| Q_2 G
|\Phi_1(-\gamma_1)\rangle = \langle\Phi_0(\gamma_1)|Q_2|\Phi_1(\gamma_1)\rangle \sim 1$
($Q_2$, second phonon coordinate, is a short-hand for $(b_2^++b_2)$); 
other
terms contain a small ``overlapping'' factor
$\varepsilon \sim \exp(-2\gamma_1^2)$.
 Using rough estimations (the same as those leading to (\ref{13})) we get
 following  expression for the ``overlapping'' part of energy in the ERA:

\begin{equation}
E_{l} \simeq -2\eta_2(\beta e^{2r_2}\cosh\lambda +\alpha
 e^{2r_1} \sinh\lambda) \sim -2\eta_2\beta\, ,
\label{20}
\end{equation}
(this expression is valid for the region $\chi<1$).
In (\ref{20}), it is the $\beta$-term which yields the
 main contribution.

As for the SRA, its counterpart reads as $\langle\Phi_0(\gamma_1)|Q_2
|\Phi_0 (-\gamma_1)\rangle$
and vanishes due to symmetry, leaving us only smaller terms
$\propto\varepsilon$,:

\begin{equation}
E_{l}\simeq -4\beta \eta_1 \varepsilon  \,
\label{21}
\end{equation}
 (the higher order terms $\sim \varepsilon^2$, $\varepsilon^2\eta$
etc. are omitted).

Comparing these expressions we can see that ERA
 yields better use of the reflection symmetry property of
Hamiltonian contributing greatly to the overlapping integral (to the
negative exchange energy)
while for the SRA this principal contribution
vanishes merely because of symmetry, leaving us with minor
contributions $\sim \varepsilon$ only,
thus there the whole idea of the reflective
ansatz losses much of its effectivity. 

This effect of lowering energy in the excitation ansatz due to overlapping
finds its origin in the presence of phonon-$2$ assistance. In the dimer or
exciton models instead of $\beta\hat{Q_2}\sigma_x$ term of the model Hamiltonian
(\ref{1}) there stands merely $\Delta\sigma_x$ with the constant
$\Delta$\cite{Wagner:1994} and this
principal part of the exchange energy (overlapping integral) for the excitation 
ansatz vanishes.


The very general impression from Figs 3-5 (b) makes us to state that introducing 
nonunitary parameters ($\eta_i$) essentially improves the variational treatment for the 
self-trapping region; In the tunnelling region merely Gaussian
expressions for displaced oscillators with squeezing (parameters $r_i$)  
gives us a satisfactory fit. Indeed, as it was demonstrated from analytical estimations 
and as it is seen from Figs. 1 a,c, the admixture of reflection part is relevant rather
for selftrapping region where this form of the trial function is the best
choice.

However, the closer we are to the intermediate region between selftrapping and
tunnelling phases, the stronger are discrepancies
for all curves. As it is seen from Fig.1 b it is the case where the wavefunctions
display their radial symmetric structure. 
Examining on Figs. 3-5  the curves corresponding to variational ansatzes 
with or without mixing parameter $\lambda$ we see that this parameter 
essentially 
lowers the energy exactly in the region of $\chi\simeq 1$. It is worth
comparing Fig. 3 for $\mu=0.5$ (weak coupling)
with Figures 4 and 5 ($\mu=1,\mu=4$), both representing strong couplings.
We see immediately that the coupling parameter $\lambda$ gains importance 
rather for 
small couplings where it improves the results for wider range of $\chi$, and not
only for $\chi\simeq 1$.

Fig. 6 where
we plotted the differences of the variational energy calculated with and 
without taking 
into account the mode correlation $S_{12}(\lambda )$
illustrates this statement (see also (20)) by showing the regions of 
importance of the
correlation parameter $\lambda$ in the whole plane ($\chi, \mu$).
The mode correlation represented
by  $\lambda$ (Fig. 6) appears to be by one order larger than the contribution
of the competing nonlinearity due to the reflection level
mixing $\eta_1$.
The correlation $\lambda$ is most effective for weak effective couplings $\mu$ at
$\chi\simeq (0.5,1.5)$ where it competes the selflocalization in support of
the tunneling phase.
For large $\mu$ it contributes only very close to $\chi=1$, where it reveals
a maximum for all $\mu$. 
This is quite understandable if we note that introducing $S_{12}(\lambda)$
(\ref{11}) means effective rotation of trial functions (displaced Gaussians)
in the plane $Q_1\otimes Q_2$, and $\phi^{+}(\lambda)$, $\phi^{-}(-\lambda)$ are
rotated in the opposite directions symmetrically with respect to the line
$Q_1=0$, which indeed repeats Fig.1b, fitting the rotation symmetry features.
This degree of freedom allows to represent
the picture of the wave function especially in the transition region where
selftrapping and tunnelling regions are mixed together and are hardly
distinguishable, which is the case of weak couplings. At strong couplings
those regions are more pronounced, the border between them is sharper;
in this case the parameter $\lambda$ looses its importance with exception
of the vicinity of $\chi\simeq 1$ (Fig. 6).

In the case of omitting $\lambda$, from (\ref{B1})-(\ref{B5})  one can see that the
optimized value $r_2= 0$, i.e. the contribution
due to $r_2$ in (\ref{B1}) is mediated merely by the correlation $\lambda$.
For $\lambda\neq 0$, the squeezing $r_2$
significantly interplays  with $r_1$ especially for small $\mu$ and
$\chi\approx 1$, as it brings almost half of the contribution of $r_2$.
This effect was omitted in the variational treatment of E$\otimes$e model
by Lo\cite{Lo:1991a}. Some authors disregarded this circumstance
setting $r_1=r_2$ for simplicity, but omitting the mode correlation
($\lambda$) which is therefore not selfconsistent.

In the lattice case, the coupling with the lattice is represented by a
 transfer term in the Hamiltonian of the order of magnitude of the bandwidth
 $T$\cite{Majernikova:2002}. When $T$ is sufficiently large so that 
the ``transfer''
part of the energy is comparable to the ``local'' energy contribution, the 
effect of $\eta_1$ is considerably stronger, and we do not observe
suppression caused by introducing extra correlation between phonon modes 
(that is an analogue to the parameter $\lambda$). The phonon-$1,2$
correlation in the lattice case is of smaller order of magnitude than
the contributions from the transfer terms (respectively of the order
$\beta \exp(-\gamma _1^2)$ and $T$). Because of that introducing the
correlation VP in the lattice case does not yield considerable improvement
of the results.

 \section{Conclusion}

Hamiltonian (\ref{3}) allows us to distinguish two competing regimes of
the electron-phonon system according to the relations of the parameters
specifying (a) selflocalization $\alpha$ vs quantum fluctuations $\Omega$
and (b) tunneling $\beta$ vs selflocalization $\alpha$.
Then, in terms of the relevant parameters
$\mu=\frac{\alpha^2}{2\Omega^2}$ (effective interaction) and
$\chi=\frac{\beta}{\alpha}$ (asymmetry), two quantum regions can be identified:
(i) $\mu \leq 0.5 $ and (ii) $\beta/\alpha \geq 1$.
In these regions the quantum fluctuations are most pronounced and
variational ansatz which pretends to be the most suitable should fit
there the numerical data at best. Our choice of the wave function
(\ref{7})-(\ref{11}) covers both regions in a complementary way:
while the choice of ERA with {\it the admixture of the excited state of the
symmetric phonon mode weighted by $\eta_2$ and including the mode
correlation $\lambda $
improves greatly the variational results in the selftrapping region},
$\chi \leq 1$, in the whole
range of $\mu$ (Figs. 2, 4, and 5), with increasing $\mu $ effectiveness of
all quantum variational parameters vanishes except of $\eta_2$
(Figs. 4 and 5).
In the tunneling region, $\chi>1$, the choice of ERA looses its
justification in benefit of SRA,  while all the remaining parameters
keep their effectiveness even for large $\mu$ (Figs. 4 and 5).

The cooperative effect of the reflection
(antisymmetric phonon mode) and of the assistance of the symmetric mode in
the tunneling results in a nonlinear interplay of both modes. It
consists in {\it the competition between the negative contribution of the
overlapping of the wave functions of different parity with respect to the
reflection and the increase of excitation energy} of the respective
 mode. This concept leads to the effective
energy decrease of the excited symmetric
reflected mode (ERA) rather than of its ground state (SRA).

The complex nonlinear interplay of the modes was elucidated by exact
numerical diagonalization of Hamiltonian (\ref{3}): in fact, we took
inspiration for  ERA from the exact ground state
wave function for $\alpha \geq \beta$ (Fig. 1a).
It suggested presence of admixture of
the first excited oscillator state of the symmetric mode in the reflected
part of the wave function. In the case of $\alpha
<\beta $,  the numerical wave function exhibited only a single symmetric peak
of a well defined harmonic oscillator.
The peak was located close to the center of the reflection symmetry
$Q_1=0$ but displaced by the phonons $2$. It corresponded to a {\it new
minimum of the effective potential which opened due to the
"bond selflocalization" on account of the joint effect of both modes} (Fig. 2c).
Note that these states are well localized, in the contrast to the states in the
``selftrapping'' region. 
It is interesting to mention in this context a special 
class of states in the excited spectra of J-T models which were
called ``exotic states'' \cite{Wagner:1989,Wagner:1992} and
were characterized by a pronounced localization of the corresponding phonon
mode. These states in the excitation spectra can be explained as the
consequence of the energy resonance and hence the tunnelling between two wells
of the effective
potential (visualized, e.g., on Fig. 2) which suggested opening of the
additional potential well in the position $\gamma_1\simeq 0$ where the exotic 
mode is to be localized. 
Although in the present paper we investigated merely the ground state
of the model, we can see that our {\it localized modes in the tunnelling dominated
region ($\chi >1$) have essentially the same origin (the ``localized'' minimum 
at zero displacement along $Q_1$ of the effective potential)}
and bear interesting resemblance to the Wagner's exotic states.
In support of this statement the full
spectrum of the phonon states, and not the ground one only
should be examined, but this problem merits a special paper. We just mention 
now our own results on numerical calculation of the energy spectrum for
the excited modes in the tunnelling region which show specific periodic 
(in model parameter $\mu$) chaotic 
``windows'' especially for higher 
modes. This periodicity in the coupling strength, which we at the 
very beginning scaled by the phonon frequency $\Omega$ clearly indicated 
the resonance behaviour, i.e. occasional coincidence of two incommensurable 
characteristic frequencies of the system whose nature can be identified with the origin of
Wagner's resonant exotic states.
For low lying modes like those of the ground state this behaviour is not so
pronounced, but merely its traces are also recognizable.

 The results exposed above, especially when speaking about the validity
of the variational approach chosen, are most adequate far from
the rotationally symmetric Jahn-Teller case ($\alpha=\beta$, or $\chi=1$) which  
was to be expected on basis of the comment to the FG transformation in the
Section II. In this case one gets an almost degenerated
degree of freedom in the space of variational parameters; namely, if
we introduce an analogue of the polar coordinates in the
 phonon-1,2 space (as some authors\cite{Obrien:1993}$^{,}$
 \cite{Wagner:1992}$^{,}$ \cite{Barentzen:2001} do),
 certain degeneracy of the energy profile over the
 angular coordinate would be observed  and
 thus this angular parameter should have been excluded from being variated.
(Strictly speaking, the angularly degenerated true ``mexican hat'' appears only 
in adiabatic approximation\cite{Obrien:1993}).
The authors exploiting various variational treatments disregarding this
circumstance must have encountered such problems for the E$\otimes$e Jahn-Teller
case. However, these inconsistencies 
become crucial rather at big coupling strength $\mu$, e.g., Fig. 4, 5).
However, the rotational symmetry can be to some extent retained even within 
the formalism of rectangular $Q_1\otimes Q_2$ space
by introducing the mode mixing parameter $\lambda$ which gains its 
significance especially in the region of weak coupling effectively 
spanning both regions (Fig. 6).
The variational approach exploiting essentially
the rotational symmetry of the model should present a complementary 
description suitable in the vicinity of $\chi=1$,
the whole problem however being a subject of our further considerations.


The support from the Grant Agency of the Czech Republic of our
project No. 202/01/1450 is highly acknowledged.
 We thank also
the grant agency VEGA (No. 2/7174/20) for partial support.


\section*{Appendix A. }

We have used following formulas for $D_i$, $S_i$, and $S_{12}$
defined by (\ref{9})-(\ref{11}) which can be found elsewhere
\cite{Kral:1990}$^{,}$\cite{Bonny:1986}:

\begin{eqnarray}
D_i(\gamma_i)^{-1}b_i D_i(\gamma_i)=b_i+\gamma_i,\ i=1,2,\label{22aa}\\
D_{1}(\gamma_1)S_1(r_1)= S_{1}(r_1)D_{1}(\tilde \gamma_1), \
\tilde\gamma_1=\gamma_1 e^{-2r},
\label{22a}\\
S_i^{-1}(r_i)b_i S_i(r_i) = b_i\cosh 2r_i+ b_i^{\dag} \sinh 2r_i,
\label{23a}\\
S_{12}^{-1}(\lambda) b_1 S_{12}(\lambda)= b_1\cosh \lambda- b_2^{\dag}
\sinh \lambda,
\label{24}\\
\langle 0|S_1^{\dag}(r_1)D_{1}^{\dag}(\gamma_1)\exp(\lambda
b_1^{\dag})\rangle= \frac{1}{(\cosh 2r_1)^{1/2}}\times \nonumber\\
\times\exp \left [\frac{\lambda^2}{2}\tanh (2r) -\lambda\gamma_1
 (\tanh 2r -1)+\frac{\gamma_1^2}{2} (\tanh 2r -1) \right ]
\label{25}, \\
\langle 0|S_1(r)D_{1}(\gamma_1)b_1^{\dag m}
\rangle=\frac{d^m}{d\lambda^m}
\langle 0|S_1(r_1)D_{1}(\gamma_1 )\exp(\lambda
b_1^{\dag})\rangle  |_{\lambda=0},
\label{26}\\
 S_{12}(\lambda)= T^{\dag}\left (\frac{\pi}{4} \right )S_1
 \left (\frac{\lambda}{2}\right )
 S_2 \left (-\frac{\lambda}{2}\right )T\left (\frac{\pi}{4} \right ),
\label{27}\\
T(\delta) =\exp \left (\delta(b_1^{\dag}b_2-b_1 b_2^{\dag})\right )\\
T\left(\frac{\pi}{4}\right )\left (\matrix {b_1\cr b_2}\right )
T^{\dag}\left(\frac{\pi}{4}\right )=
\frac{1}{\sqrt 2} \left (\matrix {b_1-b_2\cr b_1+b_2}\right ),
\label{28}\\
T^{\dag}|0\rangle =T|0\rangle =|0\rangle.
\label{29}
\end{eqnarray}

\section*{Appendix B.}

\begin{eqnarray}
 \langle H\rangle = \frac{1}{C}\left
 [\langle(\Psi_{0}^{(+)\dag}+\eta_1\Psi_{0}^{(-)\dag})H(\Psi^{(+)}_{0}+
 \eta_1\Psi^{(-)}_{0})
 \rangle +2\eta_2 \langle \Psi_{0}^{(+)\dag} H
 \Psi^{(-)}_{1}\rangle\right.\nonumber\\
 \left.+2\eta_1\eta_2
 \langle \Psi_{0}^{(-)\dag} H \Psi^{(-)}_{1}\rangle+\eta_2^2 \langle
 \Psi_{1}^{(-)\dag} H \Psi^{(-)}_{1}\rangle\right ]
 \label{B1}
 \end{eqnarray}
 where
\begin{eqnarray}
\langle(\Psi_{0}^{(+)\dag}+\eta_1\Psi_{0}^{(-)\dag})H(\Psi^{(+)}_{0}+
\eta_1\Psi^{(-)}_{0}
 \rangle = (1+\eta_1^2+2\eta_1\varepsilon )
\left [  \frac{1}{2} (\cosh 4r_1+\cosh 4r_2)
\cosh 2\lambda+\gamma _1^2+\gamma _2^2  \right]  \nonumber\\
 + 2\eta_1\varepsilon
\left\{ -\tanh 2\lambda\sinh 2\lambda\cosh 2(r_1+r_2) \cosh
2(r_1-r_2)   \right.\nonumber\\
\left.+\tilde\gamma _1^2[(e^{2(r_1+r_2)}-e^{(-2(r_1+r_2))}\cosh 4\lambda)
(1+\tanh^2 2\lambda)
 +2 e^{(-2(r_1+r_2))}\sinh 4\lambda\tanh 2\lambda]\cosh
2(r_1-r_2)\right.\nonumber\\
\left.+\frac{\tilde\gamma _1^2}{\cosh^2 2\lambda}(\sinh 4r_1-\sinh 4
r_2) -2\gamma _1(\tanh (2\lambda)
e^{2(r_2-r_1)}\gamma _2+\gamma _1)\right\}
\nonumber\\
 +\frac{2\alpha}{\Omega}
(1-\eta_1^2)\gamma_1-\frac{2\beta}{\Omega} \left [(1+\eta_1^2)[\gamma_2
-\gamma_1\tanh 2\lambda e^{2 (r_2-r_1)}]\varepsilon
+2 \eta_1 \gamma _2 \right ]
\label{B2}
 \end{eqnarray}

 \begin{eqnarray}
 \langle \Psi^{(+)}_{0}H\Psi^{(-)}_{1}\rangle=
 \left [ \frac{1}{2}(\cosh 4r_1+\cosh 4r_2)\cosh
 2\lambda
+\gamma_1^2+\gamma_2^2\right ]
\frac{1}{\sqrt 2}(\langle b_1...\rangle +\langle b_2...\rangle )\nonumber\\
+\frac{1}{2\sqrt 2}\left (
(\cosh 2(\lambda+2r_1)+\cosh 2(\lambda+ 2r_2))
 ]\langle b_2...\rangle
   + (\cosh 2(\lambda-2r_1)+\cosh 2(\lambda-2r_2))
\langle b_1...\rangle\right)\nonumber\\
+ \frac{1}{4\sqrt 2}\left[(-\sinh 2(\lambda-2r_1)-\sinh
2(\lambda-2r_2))\langle b_1^3...\rangle + (\sinh 2(\lambda+2r_1)+\sinh
2(\lambda+2r_2))\langle b_2^3...\rangle\right]\nonumber\\
+[-\sinh 4r_1+\sinh 4r_2-\frac{1}{2}(\sinh
2(\lambda-2r_1)+\sinh 2(\lambda-2r_2))]\frac{1}{2\sqrt 2}
\langle b_1^2b_2...\rangle\nonumber\\
+[-\sinh 4r_1+\sinh 4r_2+\frac{1}{2}(\sinh
2(\lambda+2r_1)-\sinh 2(\lambda+2r_2))]\frac{1}{2\sqrt 2}
\langle b_1b_2^2...\rangle\nonumber\\
+[(\gamma_1e^{2r_1}\sinh \lambda+\gamma_2 e^{2r_2}\cosh \lambda) ]
(\langle b_1b_2 ...\rangle+\varepsilon)\nonumber\\
+\frac{1}{ 2}  (-\gamma_1 e^{2r_1}+\gamma_2
e^{2r_2})e^{-\lambda}\langle b_1^2....\rangle+\frac{1}{ 2}  (\gamma_1 e^{2r_1}+\gamma_2
e^{2r_2})e^{\lambda}\langle b_2^2....\rangle   -\frac{\beta}{\Omega}\cosh\lambda e^{2r_2}
\nonumber\\
+\frac{\alpha}{\Omega} \left [ -e^{2r_1}\sinh
\lambda (\varepsilon +\langle b_1 b_2...\rangle )
 + \frac{e^{2r_1}}{2}\left(  \langle b_1^2...\rangle e^{-\lambda}-
  \langle b_2^2...\rangle e^{\lambda}\right ) -
  \sqrt 2\gamma_1 \left (\langle
  b_1...\rangle +\langle b_2...\rangle\right )  \right]
\label{B3}
\end{eqnarray}
\begin{eqnarray}
\langle\Psi^{(-)\dag}_{0}H\Psi^{(-)}_{1}\rangle =  (\sinh \lambda
\gamma_1e^{2r_1}+\cosh\lambda\gamma_2e^{2r_2})
-\frac{\alpha}{\Omega}\sinh
 \lambda e^{2r_1}-\frac{\beta}{\Omega}
\left [\sqrt 2 \gamma_2(<b_2...>+<b_1...>)\right.\nonumber\\
\left.
+\frac{1}{2}(e^{-\lambda+2r_2}<b_2^2...>+e^{\lambda+2r_2}<b_1^2...>
+\cosh \lambda e^{2r_2}<b_1b_2...>+ \cosh \lambda e^{2r_2}<...>\right ]
\label{B4}
 \end{eqnarray}

\begin{eqnarray}
\langle \Psi^{(-)\dag}_{1}H\Psi^{(-)}_{1}\rangle= \gamma_1^2+\gamma_2^2+
\cosh 2\lambda (\cosh 4r_1+\cosh 4r_2)+\sinh
2(r_1+r_2)\sinh 2 (r_2-r_1) \nonumber\\
 -\frac{\alpha}{\Omega}2\gamma_1
-\frac{\beta }{\Omega}
 \left\{ 2\gamma_2\cosh 2\lambda  \varepsilon
 +\gamma_2(<b_2^2...>-<b_1^2...>)\sinh \lambda
\right.\nonumber\\
 \left. +\frac{e^{2r_2}}{2\sqrt 2}\left [(<b_1...>e^{-\lambda}+<b_2...>
 e^{\lambda}) 4\cosh \lambda \right.\right.\nonumber\\
\left.\left.+((<b_2^3...>-<b_1^2b_2...>)e^{\lambda}+(<b_2^2b_1...>-<b_1^3...>)
e^{-\lambda})\sinh\lambda\right]\right\}
\label{B5}
\end{eqnarray}

The mean values in (\ref{B6})
\begin{eqnarray}
\langle b_1 D_1 (\sqrt 2 \tilde \gamma_1
e^{\lambda})S_1(\lambda)\rangle\langle D_2(-\sqrt 2 \tilde \gamma_1
e^{-\lambda})S_2 (-\lambda)\rangle
=
\frac{\sqrt 2 \tilde\gamma_1 e^{-\lambda}}{\cosh ^2 2\lambda} \varepsilon ;
\label{B7}
\end{eqnarray}
\begin{eqnarray}
\langle b_1^2 D_1 (\sqrt 2 \tilde \gamma_1
e^{\lambda})S_1(\lambda)\rangle\langle D_2(-\sqrt 2 \tilde \gamma_1
e^{-\lambda})S_2 (-\lambda)\rangle\nonumber\\
= \frac{1}{\cosh 2\lambda}\left ( \tanh 2\lambda +2\tilde
\gamma_1^2e^{2\lambda}(\tanh 2\lambda -1)^2\right )
\varepsilon ;
\label{B8}
\end{eqnarray}
\begin{eqnarray}
\langle b_1^3 D_1 (\sqrt 2 \tilde \gamma_1
e^{\lambda})S_1(\lambda)\rangle\langle D_2(-\sqrt 2 \tilde \gamma_1
e^{-\lambda})S_2 (-\lambda)\rangle\nonumber\\
=\frac{-\sqrt 2 \tilde \gamma_1 e^{\lambda}(\tanh 2\lambda -1)}
{\cosh 2\lambda}\left (3 \tanh 2\lambda +2\tilde \gamma_1^2 e^{2\lambda}
(\tanh 2\lambda-1)^2\right ) \varepsilon ;
\label{B9}
\end{eqnarray}
$$
\langle b_2^k...\rangle = \langle b_1^k...\rangle | _{\gamma_1\rightarrow -\gamma_1,
\lambda\rightarrow -\lambda},  \ k=1,2,3.$$


\begin{figure}[h]
\epsfxsize=15cm
\epsfbox{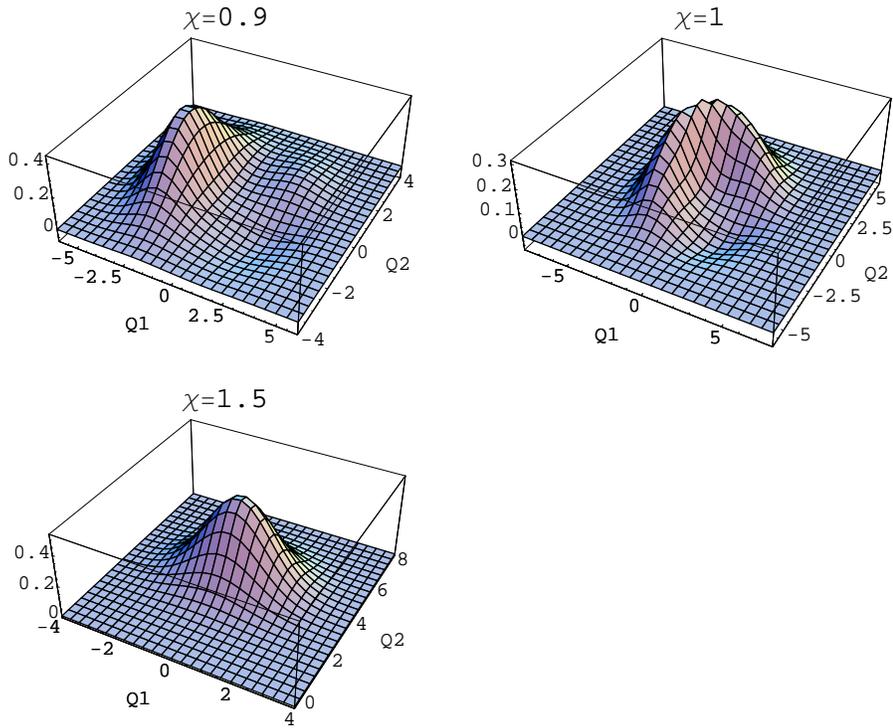}
\caption{The numerical ground state wave functions at $\mu= 2$ and
$\chi=0.9$ (a) $\chi=1$ (b) and $\chi=1.5$ (c). }
\label{fig1}
\end{figure}
\newpage

\begin{figure}[h]
\epsfxsize=10cm
\epsfbox{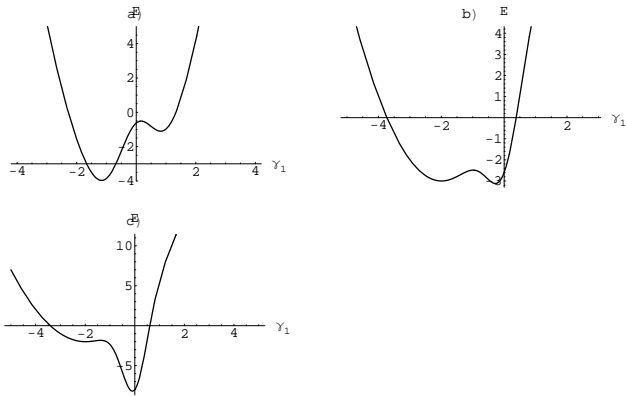}
\caption{Shapes of the effective potential corresponding to the wave functions
at Figs.1a, 1b and 1c, respectively. }
\label{fig2}
\end{figure}

\newpage

 \begin{figure}[h]
\epsfxsize=9cm
\epsfbox{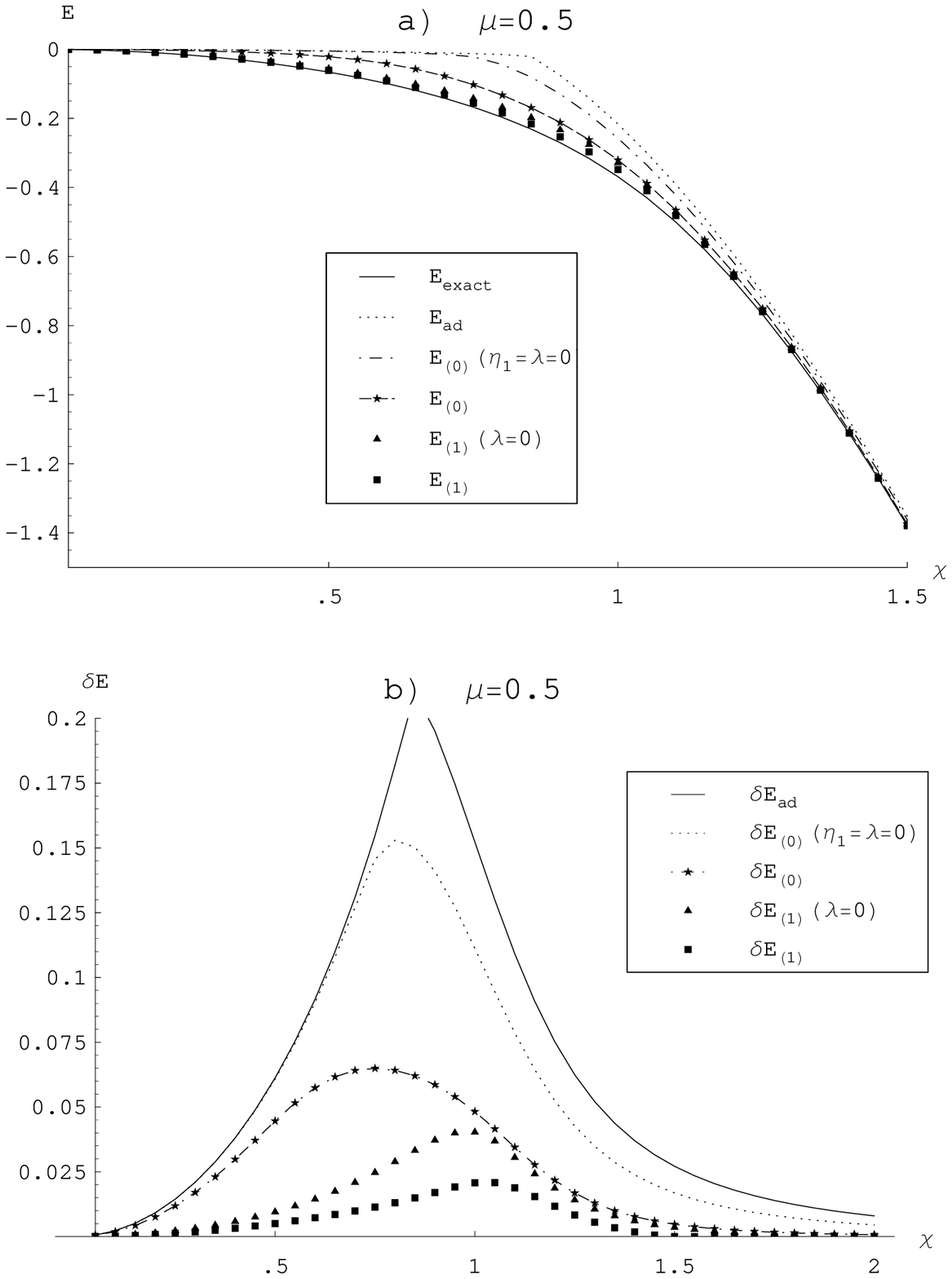}
\label{fig3}
\caption{(a) The ground state energies  (\ref{B1}) for $\mu=0.5$.
The seltrapping dominated GS spans over $\chi<1$
 and the tunneling dominated GS over $\chi>1$.
The curves plotted represent cases (from below): numerical simulation GS,
$E_{ex}$; ERA, $E_1$; ERA, $E_1(\lambda=0)$; SRA, $E_{0}$; SRA,
$E_0(\eta_1=\lambda=0)$; adiabatic  GS $E_{ad}$. \\
(b) Differences of the ground states from (a) and the exact numerical
GS.  ERA considerably improves the results for $\chi<1$. It also shifts the
maximum of the differences to the point $\chi=1$. }
\end{figure}

\newpage

 \begin{figure}[h]
\epsfxsize=9cm
\epsfbox{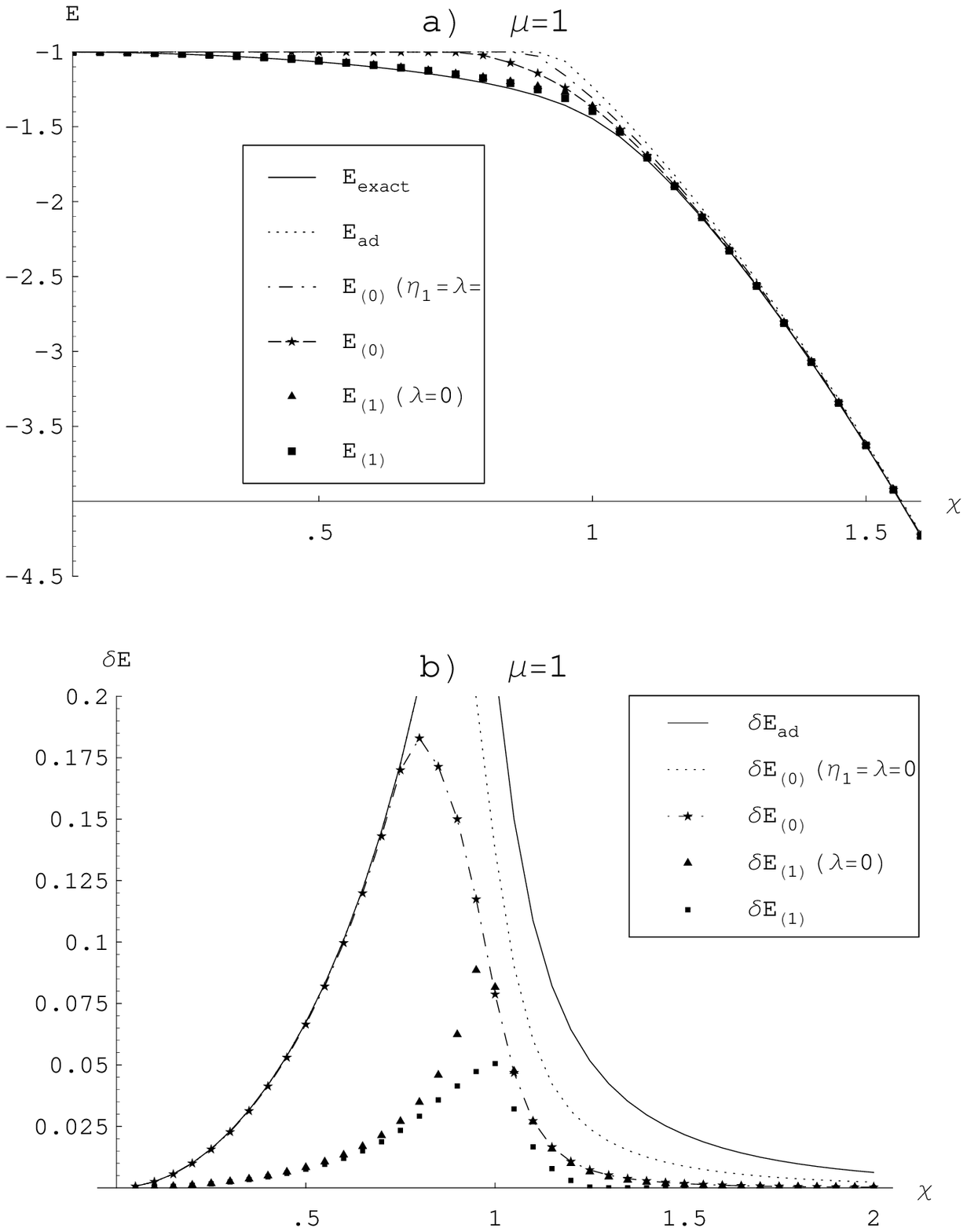}
\label{fig4}
\caption{The same as in Fig. 3 for $\mu=1$. With increasing $\mu$, for
$\chi<1$, the loss of efficiency of all VP except of $\eta_2$ is evident.}
\end{figure}

\newpage

 \begin{figure}[h]
\epsfxsize=9cm
\epsfbox{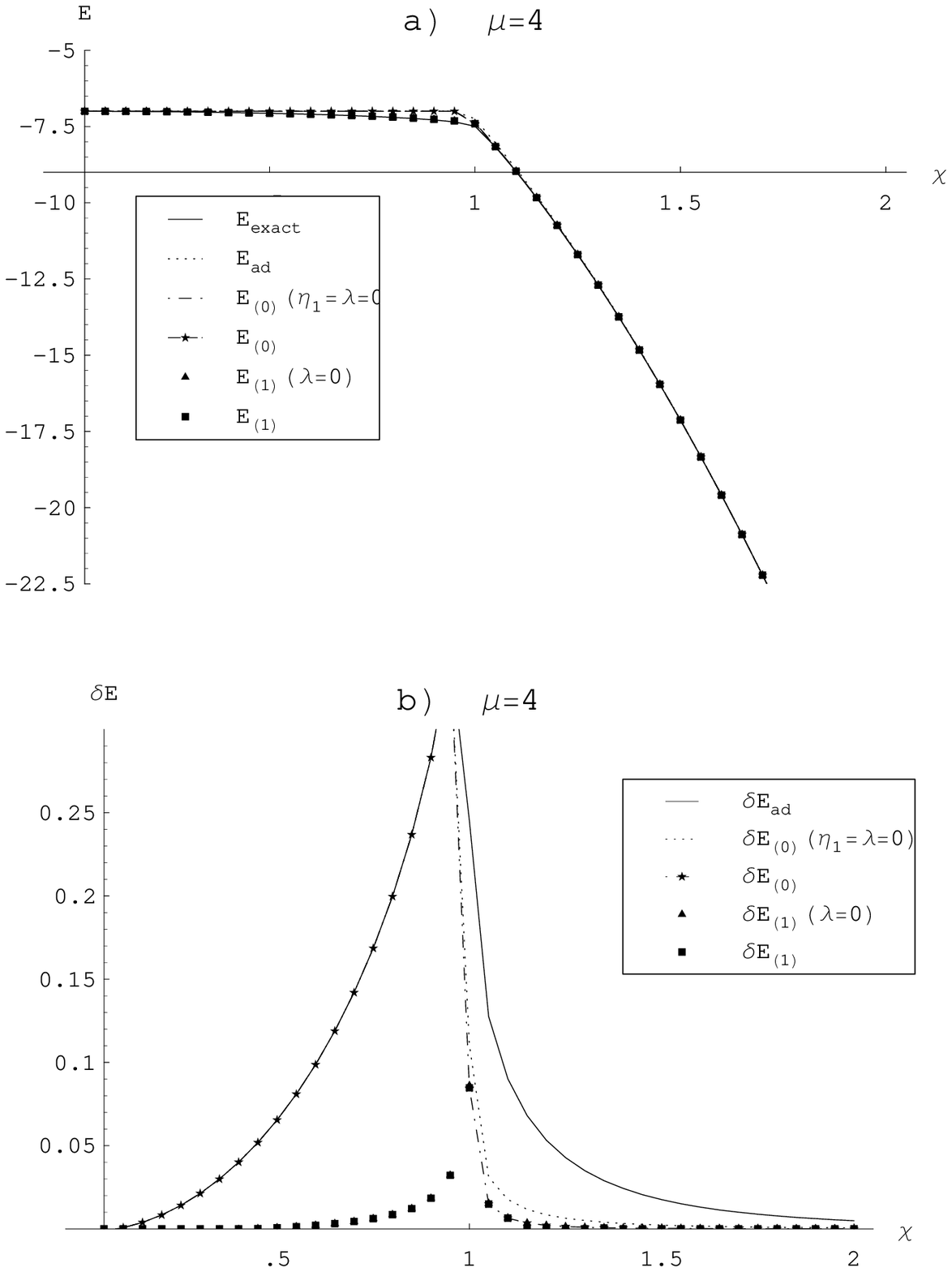}
\label{fig5}
\caption{The same as in Fig. 3 and 4 for $\mu=4$.}
\end{figure}

\newpage

 \begin{figure}[h]
\epsfxsize=9cm
\epsfbox{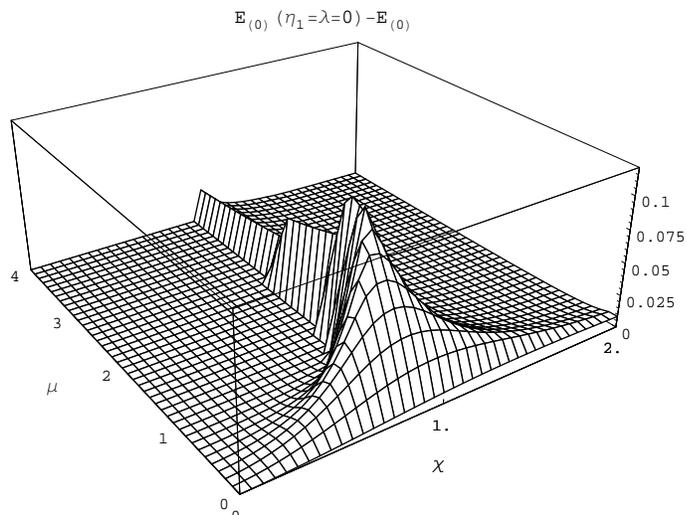}
\label{fig6}
\caption{The difference of two GS: the GS without
the reflection and correlation effects $E_0(\eta=0,\lambda=0)$ and the
GS $E_0$ including these effects.}
\end{figure}

\newpage

{\bf Figure Captions}

Fig. 1. The numerical ground state wave functions at $\mu= 2$ and
$\chi=0.9$ (a) $\chi=1$ (b) and $\chi=1.5$ (c).

Fig. 2. Shapes of the effective potential corresponding to the wave functions
at Figs.1a, 1b and 1c, respectively.

Fig. 3. 
(a) The ground state energies  (\ref{B1}) for $\mu=0.5$.
The seltrapping dominated GS spans over $\chi<1$
 and the tunneling dominated GS over $\chi>1$.
The curves plotted represent cases (from below): numerical simulation GS,
$E_{ex}$; ERA, $E_1$; ERA, $E_1(\lambda=0)$; SRA, $E_{0}$; SRA,
$E_0(\eta_1=\lambda=0)$; adiabatic  GS $E_{ad}$. \\
(b) Differences of the ground states from (a) and the exact numerical
GS.  ERA considerably improves the results for $\chi<1$. It also shifts the
maximum of the differences to the point $\chi=1$.

Fig. 4. The same as in Fig. 3 for $\mu=1$. With increasing $\mu$, for
$\chi<1$, the loss of efficiency of all VP except of $\eta_2$ is evident.

Fig. 5. The same as in Fig. 3 and 4 for $\mu=4$.
 
Fig. 6. The difference of two GS: the GS without
the reflection and correlation effects $E_0(\eta=0,\lambda=0)$ and the
GS $E_0$ including these effects.

 \end{document}